\documentclass[12pt]{article}

\newcommand{\be}{\begin{equation}}
\newcommand{\ee}{\end{equation}}
\newcommand{\bi}[1]{\vspace{-3mm} \bibitem{#1}}
\usepackage{epsfig,amsmath,amssymb,graphics,graphicx}

\begin{document}

\begin{center}
Journal of Physics A 42 (2009) 465102
\end{center}

\begin{center}

{\Large \bf  Differential Equations with Fractional Derivative \\ 
and Universal Map with Memory}
\vskip 5 mm

{\large \bf Vasily E. Tarasov$^{1,2}$}\\

\vskip 3mm

{\it $1)$ Courant Institute of Mathematical Sciences, New York University \\
251 Mercer Street, New York, NY 10012, USA }\\ 
{\it $2)$ Skobeltsyn Institute of Nuclear Physics, \\
Moscow State University, Moscow 119991, Russia } \\
{E-mail: tarasov@theory.sinp.msu.ru}

\vskip 11 mm

\begin{abstract}
Discrete maps with long-term memory are obtained 
from nonlinear differential equations 
with Riemann-Liouville and Caputo fractional derivatives.
These maps are generalizations of the well-known universal map.
The memory means that their present state is determined 
by all past states with special forms of weights. 
To obtain discrete map from fractional differential equations, 
we use the equivalence of the Cauchy-type problems and
to the nonlinear Volterra integral equations of second kind.
General forms of the universal maps with memory,
which take into account general initial conditions,
for the cases of the Riemann-Liouville and Caputo fractional derivatives, 
are suggested.
\end{abstract}

\end{center}

\section{Introduction}

A dynamical system consists of a set of possible states, together
with a rule that determines the present state in terms of past states.
If we require that the rule be deterministic, then 
we can define the present state uniquely from the past states.
A discrete-time system without memory
takes the current state as input and 
updates the situation by producing a new state as output.
All physical classical models are described by differential or 
integro-differential equations, and 
the discrete-time systems can be considered as 
a simplified version of these equations.
A discrete form of the time evolution equation is called the map.
Maps are important because they encode the behavior of deterministic systems.
The assumption of determinism is that the output of the map 
can be uniquely determined from the input.
In general, the present state is uniquely determined by all past states, 
and we have a discrete map with memory.
Discrete maps are used for the study of evolution problems, 
possibly as a substitute of differential equations \cite{Chirikov,SUZ,Zaslavsky1}. 
They lead to a much simpler formalism, which is particularly useful in simulations.
The universal discrete map is one of the most widely studied maps.
In this paper, we consider discrete maps with memory
that can be used to study solutions of 
fractional differential equations \cite{SKM,MR,Podlubny,KST}. 

The nonlinear dynamics can be considered in terms of discrete maps. 
It is a very important step in understanding the qualitative
behavior of systems described by differential equations.
The derivatives of non-integer orders are a generalization of 
the ordinary differentiation of integer order. 
Fractional differentiation with respect to time is
characterized by long-term memory effects. 
The discrete maps with memory are considered in 
\cite{Ful,Fick1,Fick2,Giona,Gallas,Stan,JPA2008}. 
The interesting question is a connection of 
fractional differential equations 
and discrete maps with memory.
It is important to derive discrete maps with memory 
from the equation of motion. 

In Ref. \cite{JPA2008}, we prove that the discrete maps with memory 
can be derived from differential equations with fractional derivatives.
The fractional generalization of the universal map 
was obtained \cite{JPA2008} from a differential equation 
with Riemann-Liouville fractional derivatives.
The Riemann-Liouville derivative has some notable 
disadvantages such as the hyper-singular improper integral, 
where the order of singularity is higher than the dimension, and 
nonzero of the fractional derivative of constants, 
which would entail that dissipation does not vanish for a 
system in equilibrium.  
The desire to formulate initial value problems for mechanical
systems leads to the use of Caputo fractional derivatives 
rather than Riemann-Liouville fractional derivative.

It is possible to state that the Caputo fractional derivatives allows us 
to give more clear mechanical interpretation.
At the same time, we cannot state that the Riemann-Liouville 
fractional derivative does not have a physical interpretation 
and that it shows unphysical behavior.
Physical interpretations of the Riemann-Liouville fractional derivatives 
are more complicated than Caputo fractional derivatives. 
But the Riemann-Liouville fractional derivatives naturally 
appear for real physical systems in electrodynamics.
We note that the dielectric susceptibility of a wide class of 
dielectric materials follows, over extended frequency ranges, 
a fractional power-law frequency dependence that is called
the "universal" response \cite{Jo2,Jrev}. 
As was proved in \cite{JPCM2008-2,JPCM2008-1}, 
the electromagnetic fields in such dielectric media 
are described by differential equations  
with Riemann-Liouville fractional time derivatives.
These fractional equations for "universal" electromagnetic waves 
in dielectric media are common to a wide class of materials, 
regardless of the type of physical structure, 
chemical composition, or of the nature of the polarizing species.
Therefore, we cannot state that Riemann-Liouville fractional time derivatives
do not have a physical interpretation.
The physical interpretation of these derivatives
in electrodynamics is connected with
the frequency dependence of the dielectric susceptibility.
As a result, the discrete maps with memory that are connected with 
differential equations with 
Riemann-Liouville fractional derivatives are very important,
and these derivatives naturally appear for real physical systems.

For computer simulation and physical application, 
it is very important to take into account the initial conditions 
for discrete maps with memory that are obtained from 
differential equations with Riemann-Liouville fractional time derivatives.
In Ref. \cite{JPA2008}, these conditions are not obtained.
In this paper, to take into account the initial condition, 
we use the equivalence of the differential equation 
with Riemann-Liouville and Caputo fractional derivatives and 
the Volterra integral equation.
This approach is more general than the auxiliary variable method
that is used in Ref. \cite{JPA2008}.
The proof of the result for Riemann-Liouville fractional derivatives 
is more complicated in comparison with 
the results for the Caputo fractional derivative.
In this paper, we prove that the discrete maps with memory can be obtained  
from differential equations with the Caputo fractional derivative.
The fractional generalization of the universal map 
is derived from a fractional differential equation 
with Caputo derivatives.

The universal maps with memory are obtained by using the equivalence
of the fractional differential equation and the Volterra integral equation.
We reduce the Cauchy-type problem for the differential equations 
with the Caputo and Riemann-Liouville  fractional derivatives  
to nonlinear Volterra integral equations of second kind.
The equivalence of this Cauchy type problem 
for the fractional equations with the Caputo derivative 
and the correspondent Volterra integral equation 
was proved by Kilbas and Marzan in \cite{KM1,KM2}.  
We also use that the Cauchy-type problem for the differential equations 
with the Riemann-Liouville fractional derivative can be reduced 
to a Volterra integral equation. 
The equivalence of this Cauchy-type problem 
and the correspondent Volterra equation 
was proved by Kilbas, Bonilla, and Trujillo in \cite{KBT1,KBT2}.

%%%%%%%%%%%%%%%%%%%%%%%%%%%%%%%%%%%%%%%%%%%%%%%%%%%%%%%%%%%%%%%%%%55

In Section 2, differential equations with integer derivative and 
universal maps without memory are considered 
to fix notations and provide convenient references. 
In Section 3, fractional differential equations 
with the Riemann-Liouville derivative 
and universal maps with memory are discussed. 
In Section 4, the difference between 
the Caputo and Riemann-Liouville fractional derivatives are discussed.
In Section 5, fractional differential equations with the Caputo derivative 
and correspondent discrete maps with memory are considered. 
A fractional generalization of the universal map is obtained from
kicked differential equations with the Caputo fractional 
derivative of order $1< \alpha \le 2$.
The usual universal map is a special case of the 
universal map with memory. 
Finally, a short conclusion is given in section 6.

%%%%%%%%%%%%%%%%%%%%%%%%%%%%%%%%%%%%%%%%%%%%%%%%%%%%%%%%%%%%%%%%%%%%%%%%%%

\section{Integer derivative and universal map without memory}

In this section, differential equations with derivative of integer order 
and the universal map without memory are considered 
to fix notations and provide convenient references.

Let us consider the equation of motion
\be \label{eq1}
D^2_t x(t)+ K G[x(t)] \sum^{\infty}_{k=1} \delta \Bigl(\frac{t}{T}-k \Bigr)=0
\ee
in which perturbation is a periodic sequence of delta-function-type pulses (kicks)
following with period $T=2\pi / \nu$, $K$ is an amplitude of the pulses, 
$D^2_t=d^2/dt^2$, and $G[x]$ is some real-valued function. 
It is well-known that this differential equation 
can be represented in the form of the discrete map 
\be \label{UM}
x_{n+1}-x_n=p_{n+1}T , \quad p_{n+1}-p_n=-KT\, G[x_n] .
\ee
Equations (\ref{UM}) are called the universal map.
For details, see for example \cite{Chirikov,SUZ,Zaslavsky1}.

The traditional method of derivation of the universal map equations from
the differential equations is considered in Section 5.1 of \cite{SUZ}.
We use another method of derivation of these equations 
to fix notations and provide convenient references. 
We obtain the universal map by using the equivalence
of the differential equation and the Volterra integral equation. \\

\vskip 3mm
\noindent
{\large \bf Proposition 1.}
{\it The Cauchy-type problem for the differential equations 
\be
D^1_t x(t) =p(t) ,
\ee
\be 
D^1_t p(t)=- 
K \, G[x(t)] \sum^{\infty}_{k=1} \delta \Bigl(\frac{t}{T}- k \Bigr) , 
\ee
with the initial conditions
\be %%%\label{RL5}
x(0)=x_0, \quad p(0)=  p_0 
\ee
is equivalent to the universal map equations of the form }
\be \label{E6i}
x_{n+1}=x_0 + p_0 (n+1)T - K T^2 \sum^{n}_{k=1} G[x_k] \, (n+1-k) ,
\ee
\be \label{E7i}
p_{n+1}= p_0 - K T \sum^{n}_{k=1} G[x_k] .
\ee

\vskip 3mm
{\bf Proof.}
Consider the nonlinear differential equation of second order
\be \label{E1i}
D^2_t x(t) = G[t,x(t)] , \quad (0 \le t \le t_f)
\ee
on a finite interval $[0,t_f]$ of the real axis, with the initial conditions
\be \label{E2i}
x(0)=x_0, \quad (D^1_t x)(0)=p_0 . 
\ee
The Cauchy-type problem of the form (\ref{E1i}), (\ref{E2i}) 
is equivalent to the Volterra integral equation of second kind 
\be \label{E3i}
x(t)=x_0+p_0 t + \int^t_{0} \, d\tau \, G[\tau,x(\tau)] \, (t-\tau) .
\ee
Using the function
\[ G[t,x(t)]= - K G[x(t)] \sum^{\infty}_{k=1} \delta \Bigl(\frac{t}{T}-k \Bigr) , \]
for $nT<t<(n+1)T$, we obtain 
\be \label{E4i}
x(t)=x_0 + p_0 t - K T \sum^{n}_{k=1} G[x(kT)] \, (t-kT) .
\ee
For the momentum $p(t)=D^1_t x(t)$, equation (\ref{E4i}) gives
\be \label{E5i}
p(t)= p_0 - K T \sum^{n}_{k=1} G[x(kT)] .
\ee
The solution of the left side of the $(n+1)$th kick
\be \label{not1}
x_{n+1}=x(t_{n+1}-0)=\lim_{\varepsilon \rightarrow 0+} x(T(n+1)-\varepsilon), 
\ee
\be \label{not2}
p_{n+1}=p(t_{n+1}-0)=\lim_{\varepsilon \rightarrow 0+} p(T(n+1)-\varepsilon), 
\ee
where $t_{n+1}=(n+1)T$, has the form (\ref{E6i}) and (\ref{E7i}).  

This ends the proof. $\ \ \ \Box$ \\

{\bf Remark 1}. \\
We note that equations (\ref{E6i}) and (\ref{E7i}) 
can be rewritten in the form (\ref{UM}).   
Using equations (\ref{E6i}) and (\ref{E7i}), 
the differences $x_{n+1}-x_n$ and $p_{n+1}-p_n$ give 
equations (\ref{UM}) of the universal map.

{\bf Remark 2}. \\
If $G[x]=-x$, then equations (\ref{UM}) give the Anosov-type system
\be
x_{n+1}-x_n=p_{n+1}T , \quad p_{n+1}-p_n=KT x_n .
\ee
For $G[x]=\sin (x)$, equations (\ref{UM}) are
\be
x_{n+1}-x_n=p_{n+1}T , \quad p_{n+1}-p_n=-KT \, \sin(x_n) .
\ee
This map is known as the standard or 
Chirikov-Taylor map \cite{Chirikov}. 

%%%%%%%%%%%%%%%%%%%%%%%%%%%%%%%%%%%%%%%%%%%%%%%%%%%%%%%%%%%%%%%%%%%%%%%%

\section{Riemann-Liouville fractional derivative and universal map with memory}

In this section, we discuss nonlinear differential equations 
with the left-sided Riemann-Liouville fractional derivative
$ _0D^{\alpha}_t$ defined for $\alpha>0$ by 
\be \label{RLFD}
_0D^{\alpha}_t x(t)=D^n_t \ _0I^{n-\alpha}_t x(t)=
\frac{1}{\Gamma(n-\alpha)} \frac{d^n}{dt^n} \int^{t}_0 
\frac{x(\tau) d \tau}{(t-\tau)^{\alpha-n+1}} , \quad (n-1 <\alpha \le n) ,
\ee
where $D^n_t=d^n/dt^n$, and $ _0I^{\alpha}_t$ 
is a fractional integration \cite{SKM,KST,Podlubny}. 

We consider the fractional differential equation
\be \label{RL1}
 _0D^{\alpha}_t x(t)= G[t,x(t)] ,
\ee
where $G[t,x(t)] $ is a real-valued function, $0 \le n-1 < \alpha \le n$, and $t>0$, 
with the intial conditions
\be \label{RL2}
( \, _0D^{\alpha - k}_t x)(0+) =c_k, \quad k=1,...,n .
\ee
The notation $( \, _0D^{\alpha - k}_t x)(0+)$ means that
the limit is taken at almost all points of the right-sided
neighborhood $(0,0+\varepsilon)$, $\varepsilon >0$, of zero
as follows
\[ ( \, _0D^{\alpha - k}_t x)(0+) =\lim_{t \rightarrow 0+}
\, _0D^{\alpha - k}_t x(t) , \quad (k=1,...,n-1) , \]
\[ ( \, _0D^{\alpha - n}_t x)(0+) =\lim_{t \rightarrow 0+}
\, _0I^{n-\alpha}_t x(t) .  \]

The Cauchy-type problem (\ref{RL1}) and (\ref{RL2}) can be reduced 
to the nonlinear Volterra integral equation of second kind
\be \label{RL3}
x(t)=\sum^n_{k=1} \frac{c_k}{\Gamma(\alpha - k +1)} t^{\alpha-k} 
+ \frac{1}{\Gamma(\alpha)} \int^t_0 \frac{G[\tau,x(\tau)] d\tau}{(t-\tau)^{1-\alpha}} ,
\ee
where $t>0$.
The result was obtained by Kilbas, Bonilla, and Trujillo in \cite{KBT1,KBT2}.
For $\alpha=n=2$, equation (\ref{RL3}) gives (\ref{E3i}).

The Cauchy-type problem (\ref{RL1}) and (\ref{RL2}) and 
the nonlinear Volterra integral equation (\ref{RL3}) 
are equivalent in the sense that, if  $x(t) \in L (0,t_f)$
satisfies one of these relations, then it also satisfies the other.
In \cite{KBT1,KBT2} (see also Theorem 3.1. in Section 3.2.1 of \cite{KST}),  
this result is proved by assuming that the function $G[t,x]$ 
belongs to $L (0,t_f)$ for any $x \in W \subset \mathbb{R}$.

Let us give the basic theorem regarding the nonlinear
differential equation involving 
the Riemann-Liouville fractional derivative.

\vskip 3mm
\noindent
{\large \bf Kilbas-Bonilla-Trujillo Theorem.}
{\it Let $W$ be an open set in $\mathbb{R}$ and 
let $G[t,x]$, where $t \in (0,t_f]$ and $x \in W$, 
be a real-valued function such that $G[t,x] \in L(0,t_f)$ for any $x \in W$.
Let $x(t)$ be a Lebesgue measurable function on $(0,t_f)$.
If $x(t) \in L (0,t_f)$, then $x(t)$ satisfies 
almost everywhere equation (\ref{RL1}) and 
conditions (\ref{RL2}) if, and only if, 
$x(t)$ satisfies almost everywhere the integral equation (\ref{RL3}).} \\

{\bf Proof.}
This theorem is proved in \cite{KBT1,KBT2}  
(see also Theorem 3.1. in Section 3.2.1 of \cite{KST}). 
$\ \ \ \Box$ \\

In Ref. \cite{JPA2008} we consider a fractional generalization of 
equation (\ref{eq1}) of the form
\be \label{RL4}
_0D^{\alpha}_t x(t) + K \, G[x(t)] \sum^{\infty}_{k=1} 
\delta \Bigl(\frac{t}{T}-k \Bigr)=0,  \quad (1 <\alpha \le 2) ,
\ee
where $t>0$, and $ _0D^{\alpha}_t$ is the Riemann-Liouville 
fractional derivative defined by (\ref{RLFD}). 
Let us give the following theorem for equation (\ref{RL4}). \\

%%%\vskip 3mm
\noindent
{\large \bf Proposition 2.}
{\it The Cauchy-type problem for 
the fractional differential equation of the form (\ref{RL4})
with the initial conditions
\be \label{RL5}
( \, _0D^{\alpha-1}_t x)(0+)=c_1, \quad 
( \, _0D^{\alpha-2}_t x)(0+)=( \, _0I^{2-\alpha}_t x)(0+)=  c_2 
\ee
is equivalent to the equation 
\be \label{RL6b}
x(t)= \frac{c_1}{\Gamma(\alpha)} t^{\alpha-1} +
\frac{c_2}{\Gamma(\alpha - 1)} t^{\alpha-2} 
- \frac{K T}{\Gamma(\alpha)} \, 
\sum^{n}_{k=1}\,  \, G[x(kT)] \, (t-kT)^{\alpha-1} ,
\ee
where $nT <t< (n+1)T$.} \\

\vskip 3mm
{\bf Proof.}
Using the function
\be \label{G}
G[t,x(t)]=- K \, G[x] \sum^{\infty}_{k=1} \delta \Bigl(\frac{t}{T}-k \Bigr) , 
\ee
equation (\ref{RL4}) has the form of (\ref{RL1}) with 
the Riemann-Liouville fractional derivative of order $\alpha$, 
where $1 <\alpha \le 2$.
It allows us to use the Kilbas-Bonilla-Trujillo theorem.   
As a result, equation (\ref{RL4}) with initial conditions (\ref{RL2}) 
of the form (\ref{RL5}) 
is equivalent to the nonlinear Volterra integral equation 
\be \label{RL6}
x(t)= \frac{c_1}{\Gamma(\alpha)} t^{\alpha-1} +
\frac{c_2}{\Gamma(\alpha - 1)} t^{\alpha-2} 
- \frac{K}{\Gamma(\alpha)} \, \sum^{\infty}_{k=1}\, 
\int^t_0 \, d \tau \, G[x(\tau)] \, (t-\tau)^{\alpha-1} \, 
\delta \Bigl(\frac{\tau}{T}-k \Bigr) ,
\ee
where $t>0$.
If $nT <t< (n+1)T$, then the integration in (\ref{RL6}) 
with respect to $\tau$ gives (\ref{RL6b}).

This ends the proof. $\ \ \ \Box$ \\

To obtain equations of discrete map a momentum must be defined. 
There are two possiblities of defining the momentum:
\be
p(t)= \, _0D^{\alpha-1}_t x(t) ,\quad p(t)= D^1_t x(t) .
\ee

%%%%%%%%%%%%%%%%%%%%%%%%%%%%%%%%%%%%%%%%%%%%%%%%%%%%%%%%%%%%%%%%%%%%%%%%%%%%

Let us use the first definition. Then the momentum is defined by  
the fractional derivative of order $\alpha-1$. 
Using the definition of the Riemann-Liouville fractional derivative (\ref{RLFD}) 
in the form
\be \label{RL8}
_0D^{\alpha}_t x(t) = D^2_t \ _0I^{2-\alpha}_t x(t) , \quad (1 <\alpha \le 2) ,
\ee
we define the momentum
\be \label{RL9}
p(t)= \, _0D^{\alpha-1}_t x(t)=
\frac{1}{\Gamma(2-\alpha)} \frac{d}{dt} \int^{t}_0 
\frac{x(\tau) d \tau}{(t-\tau)^{\alpha-1}} , \quad (1 <\alpha \le 2) ,
\ee
where $x(\tau)$ is defined for $\tau \in (0,t)$.
Then
\be 
_0D^{\alpha}_t x(t) =D^1_t p(t) , \quad (1 <\alpha \le 2) .
\ee
Using momentum $p(t)$ and coordinate $x(t)$, equation  (\ref{RL4}) 
can be represented in the Hamiltonian form
\be \label{Hf1}
_0D^{\alpha-1}_t x(t)=p(t),
\ee
\be \label{Hf2}
D^1_t p(t)=- K G[x(t)] \sum^{\infty}_{k=1} \delta \Bigl(\frac{t}{T}-k \Bigr) .  
\ee

%%%%%%%%%%%%%%%%%%%%%%%%%%%%%%%%%%%%%%%%%%%%%%%%%%%%%%%%%%%%%%%%%%%%%%%%%%%%%%%

\vskip 3mm
\noindent
{\large \bf Proposition 3.}
{\it The Cauchy type problem for 
the fractional differential equations of the form (\ref{Hf1}) and (\ref{Hf2})
with the initial conditions
\be \label{RL5c}
( \, _0D^{\alpha-1}_t x)(0+)=c_1, \quad 
( \, _0D^{\alpha-2}_t x)(0+)=( \, _0I^{2-\alpha}_t x)(0+) = c_2 
\ee
is equivalent to the discrete map equations }
\be \label{FUM-RL1}
x_{n+1}= \frac{c_1 T^{\alpha-1}}{\Gamma(\alpha)} (n+1)^{\alpha-1} +
\frac{c_2 T^{\alpha-2}}{\Gamma(\alpha - 1)} (n+1)^{\alpha-2} 
- \frac{K T^{\alpha}}{\Gamma(\alpha)} \, 
\sum^{n}_{k=1} \, G[x_k] \, (n+1-k)^{\alpha-1} ,
\ee
\be \label{FUM-RL2}
p_{n+1} = c_1 - K T \, \sum^{n}_{k=1}\, G[x_k] .
\ee

\vskip 3mm
{\bf Proof.}
We use Proposition 2 to prove this statement.
To obtain an equation for the momentum (\ref{RL9}), 
we use the following fractional derivatives of power functions 
(see Section 2.1 in \cite{KST}),
\be
_aD^{\alpha}_t (t-a)^{\beta-1}= 
\frac{\Gamma(\beta)}{\Gamma(\beta - \alpha)} (t-a)^{\beta-1-\alpha} ,
\quad \alpha  \ge 0, \quad \beta >0 , \quad t>a,
\ee
\be
_0D^{\alpha}_t t^{\alpha-k}= 0 ,
\quad k=1,...,n , \quad n-1 < \alpha \le n .
\ee
These equations give
\[ _0D^{\alpha}_t t^{\alpha-1} = \Gamma(\alpha) , \]
\[ _0D^{\alpha}_t t^{\alpha-2} = 0 , \]
and 
\[ _aD^{\alpha}_t (t-a)^{\alpha-1} = \Gamma(\alpha) . \]

We note that equation (\ref{RL6b}) for $x(\tau)$ can be used 
only if $\tau \in (nT,t)$, where $nT <t< (n+1)T$.
The function $x(\tau)$ in 
the fractional derivative $_0D^{\alpha}_t$ of the form
(\ref{RL9}) must be defined for all $\tau \in (0,t)$.
We cannot take the derivative $_0D^{\alpha}_t$ of the functions
$(\tau-kT)^{\alpha-1}$ that are defined for $\tau \in (kT,t)$. 
In order to use equation (\ref{RL6b}) on the interval $(0,t)$,
we must modify the sum in equation (\ref{RL6b}) 
by using the Heaviside step function.
Then equation (\ref{RL6b}) has the form
\be \label{RL6b-new}
x(\tau)= \frac{c_1}{\Gamma(\alpha)} \tau^{\alpha-1} +
\frac{c_2}{\Gamma(\alpha - 1)} \tau^{\alpha-2} 
- \frac{K T}{\Gamma(\alpha)} \, \sum^{n}_{k=1}\,  \, 
G[x(kT)] \, (\tau-kT)^{\alpha-1} \, \theta(\tau-kT) ,
\ee
where $\tau \in (0,t)$. 
Using the relation 
\be
_0D^{\alpha}_t \Bigl( (t-a)^{\alpha-1} \, \theta(t-a)  \Bigr) =
\, _aD^{\alpha}_t (t-a)^{\alpha-1} = \Gamma(\alpha) ,
\ee
equations (\ref{RL9}) and (\ref{RL6b-new}) give
\be \label{RL10}
p(t)= c_1 -  K T \, \sum^{n}_{k=1}\, G[x(kT)] ,
\ee
where $nT<t<(n+1)T$.
Then the solution of the left side of the $(n+1)$-th kick
\be \label{RL11}
p_{n+1}= c_1 -  K T \, \sum^{n}_{k=1}\, G[x_k] .
\ee

As a result, we obtain a universal map with memory in the form of 
equations (\ref{FUM-RL1}) and (\ref{FUM-RL2}).

This ends the proof. $\ \ \ \Box$ \\

{\bf Remark 3}. \\
For $\alpha=n=2$ equations (\ref{FUM-RL1}) and (\ref{FUM-RL2})
give the usual universal map (\ref{E6i}) and (\ref{E7i}).

{\bf Remark 4}. \\
We note that the map (\ref{FUM-RL1}) and (\ref{FUM-RL2}) with 
\[ c_1=p_1, \quad c_2=0 \]
was obtained in \cite{JPA2008} in the form 
\be \label{N1-0}
x_{n+1}=\frac{T^{\alpha-1}}{\Gamma(\alpha)} 
\sum^{n}_{k=1} \, p_{k+1} V_{\alpha}(n-k+1) ,
\ee
\be \label{N2-0}
p_{n+1}=p_n-KT \, G(x_n) , \quad (1<\alpha\le 2),
\ee
where $p_1=c_1$, and the function $V_{\alpha}(z)$ is defined by
\be 
V_{\alpha}(z)=z^{\alpha-1}-(z-1)^{\alpha-1} , \quad (z\ge 1).
\ee
In Ref. \cite{JPA2008}, we obtain these map equations by using
an auxiliary variable $\xi (t)$ such that
\[ _0^CD^{2-\alpha}_t \xi(t) =x(t). \]
%%%where $ _0^CD^{2-\alpha}_t$ is the Caputo fractional derivative (\ref{CFD}).
The nonlinear Volterra integral equations 
and the general initial conditions (\ref{RL5c})  are not used in \cite{JPA2008}. 
In the general case, 
the fractional differential equation of the kicked system (\ref{RL4})
is equivalent to the discrete map equations
\be \label{N1}
x_{n+1}=\frac{T^{\alpha-1}}{\Gamma(\alpha)} 
\sum^{n}_{k=1} \, p_{k+1} V_{\alpha}(n-k+1)+
\frac{c_2 T^{\alpha-2}}{\Gamma(\alpha-1)} (n+1)^{\alpha-2} ,
\ee
\be \label{N2}
p_{n+1}=p_n-KT \, G(x_n) , \quad (1<\alpha\le 2),
\ee
where $p_1=c_1$.
Here we take into account the initial conditions (\ref{RL5c}). 
The second term of the right-hand side of equation (\ref{N1})
is not considered in \cite{JPA2008}.
Using $-1 < \alpha -2< 0$, we have
\[ \lim_{n \rightarrow \infty} (n+1)^{\alpha-2} =0 . \]
Therefore, the case of large values of $n$ is equivalent to $c_2=0$. \\

Let us give the proposition regarding the second definition
of the momentum $p(t)= D^1_t x(t)$. \\

\vskip 3mm
\noindent
{\large \bf Proposition 4.}
{\it The Cauchy-type problem for the fractional differential equations 
\be
D^1_t x(t) =p(t) ,
\ee
\be 
_0D^{\alpha}_t x(t)=- 
K \, G[x(t)] \sum^{\infty}_{k=1} \delta \Bigl(\frac{t}{T}- k \Bigr) , 
\quad (1 <\alpha < 2) ,
\ee
with the initial conditions
\be %%%\label{RL5}
( \, _0D^{\alpha-1}_t x)(0+)=c_1, \quad 
( \, _0D^{\alpha-2}_t x)(0+)=( \, _0I^{2-\alpha}_t x)(0+)=  c_2 
\ee
is equivalent to the discrete map equations }
\be \label{RL7}
x_{n+1}= \frac{c_1 T^{\alpha-1}}{\Gamma(\alpha)} (n+1)^{\alpha-1} +
\frac{c_2 T^{\alpha-2}}{\Gamma(\alpha - 1)} (n+1)^{\alpha-2} 
- \frac{K T^{\alpha}}{\Gamma(\alpha)}  \, 
\sum^{n}_{k=1} \, G[x_k]  (n+1-k)^{\alpha-1} .
\ee
\be \label{RL6d}
p_{n+1}= \frac{c_1 T^{\alpha-2}}{\Gamma(\alpha-1)} (n+1)^{\alpha-2} +
\frac{c_2 (\alpha-2) T^{\alpha-3}}{\Gamma(\alpha - 1)} (n+1)^{\alpha-3} 
- \frac{K T^{\alpha-1}}{\Gamma(\alpha-1)} \, 
\sum^{n}_{k=1} \, G[x_k]  (n+1-k)^{\alpha-2} .
\ee

\vskip 3mm
{\bf Proof.}
We define the momentum
\[ p(t)=D^1_t x(t) . \]
If $nT <t< (n+1)T$, then the differentiation of (\ref{RL6b}) 
with respect to $t$ gives
\be \label{RL6c}
p(t)= \frac{c_1}{\Gamma(\alpha-1)} t^{\alpha-2} +
\frac{c_2 (\alpha-2)}{\Gamma(\alpha - 1)} t^{\alpha-3} 
- \frac{K T}{\Gamma(\alpha-1)} \, 
\sum^{n}_{k=1}\,  \, G[x(kT)] \, (t-kT)^{\alpha-2} .
\ee
Here we use the relation
\[ \Gamma(\alpha)=(\alpha-1) \Gamma(\alpha - 1) , \quad (1<\alpha \le 2). \]
Using equations (\ref{RL6b}) and (\ref{RL6c}), we can obtain the solution 
of the left side of the $(n+1)$-th kick (\ref{not1}) and (\ref{not2}).
As a result, we have equations (\ref{RL7}) and (\ref{RL6d}).

This ends the proof. $\ \ \ \Box$ \\

{\bf Remark 5}. \\
Equations (\ref{RL7}) and (\ref{RL6d}) describe 
a generalization of equations (\ref{E6i}) and (\ref{E7i}).
If $\alpha=n=2$, and $c_2=x_0$, $c_1=p_0$, 
then equation (\ref{RL7}) gives (\ref{E6i}) and (\ref{E7i}).

{\bf Remark 6}. \\
In equations (\ref{RL6d}) and (\ref{RL6c}), we can use 
\[ \frac{c_2 (\alpha-2)}{\Gamma(\alpha - 1)}=
\frac{c_2}{\Gamma(\alpha - 2)} \]
for $1<\alpha<2$.

{\bf Remark 7}. \\
If we use the definition $p(t)=D^1_t x(t)$, then the Hamiltonian form
of the equations of motion will be more complicated 
than (\ref{Hf1}) and (\ref{Hf2}) since
\[ D^2_t \ _0I^{2-\alpha}_t x(t) \not= \ _0I^{2-\alpha}_t D^2_t x(t) . \]

{\bf Remark 8}. \\
Note that we use the usual momentum $p(t)=D^1_t x(t)$.
In this case, the values $c_1$ and $c_2$ are not connected with
$p(0)$ and $x(0)$.
If we use the momentum $p(t) =\, _0D^{\alpha-1}_t x(t)$, then $c_1=p(0)$.

%%%%%%%%%%%%%%%%%%%%%%%%%%%%%%%%%%%%%%%%%%%%%%%%%%%%%%%%%%%%%%%%%%%%%%%%%%%%%
\section{Riemann-Liouville and Caputo fractional derivatives}

In Ref. \cite{JPA2008} we consider nonlinear differential equations
with Riemann-Liouville fractional derivatives.
The discrete maps with memory  are obtained from these equations. 
The problems with initial conditions 
for the Riemann-Liouville fractional derivative are not discussed.

The Riemann-Liouville fractional derivative has some notable 
disadvantages in applications in mechanics such as 
the hyper-singular improper integral, where the order 
of singularity is higher than the dimension, and 
nonzero of the fractional derivative of constants, 
which would entail that dissipation 
does not vanish for a system in equilibrium.  
The desire to use  the usual initial value problems for mechanical 
systems leads to the use of Caputo fractional derivatives \cite{KST,Podlubny}
rather than the Riemann-Liouville fractional derivatives.

%%%%%%%%%%%%%%%%%%%%%%%%%%%%%%%%%%%%%%%%%%%%%%%%

The left-sided Caputo fractional derivative \cite{Caputo,Caputo2,GM,KST}
of order $\alpha >0$ is defined by
\be \label{Caputo}
\,  _0^CD^{\alpha}_t f(t)=
\frac{1}{\Gamma(n-\alpha)} \int^t_0 
\frac{ d\tau \, D^n_{\tau}f(\tau)}{(t-\tau)^{\alpha-n+1}} =
\, _0I^{n-\alpha}_t D^n_t f(t) ,
\ee
where $n-1 < \alpha <n$, and $_0I^{\alpha}_t$ is 
the left-sided Riemann-Liouville fractional integral 
of order $\alpha >0$ that is defined by
\be
_0I^{\alpha}_t f(t)=\frac{1}{\Gamma(\alpha)} 
\int^t_0 \frac{f(\tau) d \tau}{(t-\tau)^{1-\alpha}} , \quad (t>0).
\ee

This definition is, of course, more restrictive than 
the Riemann-Liouville fractional derivative \cite{SKM,KST} 
in that it requires the absolute integrability of the derivative
of order $n$. 
The Caputo fractional derivative first computes an ordinary
derivative followed by a fractional integral to achieve the
desire order of fractional derivative.
The Riemann-Liouville fractional derivative 
is computed in the reverse order. 
Integration by part of (\ref{Caputo}) will lead to 
\be \label{C-RL}
_0^CD^{\alpha}_t x(t)= \, _0D^{\alpha}_t x(t)-
\sum^{n-1}_{k=0}
\frac{t^{k-\alpha}}{\Gamma(k-\alpha+1)} x^{(k)}(0) .
\ee 
It is observed that the second term in equation (\ref{C-RL}) regularizes 
the Caputo fractional derivative to avoid the potentially divergence 
from singular integration at $t=0$. In addition, the 
Caputo fractional differentiation of a constant results in zero
\[ _0^CD^{\alpha}_t C=0 . \]
Note that the Riemann-Liouville fractional derivative of a 
constant need not be zero, and we have 
\[ _0D^{\alpha}_t C=\frac{t^{-\alpha}}{\Gamma(1-\alpha)} C . \]

If the Caputo fractional derivative is used instead of the 
Riemann-Liouville fractional derivative, then 
the initial conditions for fractional dynamical systems 
are the same as those for the usual dynamical systems.  
The Caputo formulation of fractional calculus can be more 
applicable in mechanics than the Riemann-Liouville formulation.

%%%%%%%%%%%%%%%%%%%%%%%%%%%%%%%%%%%%%%%%%%%%%%%%%%%%%%%%%%%%%%%%%%%%%%

\section{Caputo fractional derivative and universal map with memory}

In this section, we study a generalization of differential equation 
(\ref{eq1}) by the Caputo fractional derivative. 
The universal map with memory is derived from this fractional equation.

We consider the nonlinear differential equation of order $\alpha$, 
where $0 \le n-1 < \alpha \le n$,
\be \label{E1}
\,  _0^CD^{\alpha} x(t) = G[t,x(t)] , \quad (0 \le t \le t_f) ,
\ee
involving the Caputo fractional derivative $ _0^CD^{\alpha}_t$ 
on a finite interval $[0,t_f]$ of the real axis, 
with the initial conditions
\be \label{E2}
(D^k_t x)(0)=c_k , \quad k=0,...,n-1. 
\ee
Kilbas and Marzan \cite{KM1,KM2} proved the equivalence
of the Cauchy-type problem of the form (\ref{E1}), (\ref{E2}) and 
the Volterra integral equation of second kind 
\be \label{E3}
x(t)=\sum^{n-1}_{k=0} \frac{c_k}{k!} t^k + 
\frac{1}{\Gamma (\alpha)} \int^t_{0} d \tau \,
G[\tau,x(\tau)] \, (t-\tau)^{\alpha-1}
\ee
in the space $C^{n-1}[0,t_f]$.
For $\alpha=n=2$ equation (\ref{E3}) gives (\ref{E3i}).

Let us give the basic theorem regarding the nonlinear
differential equation involving 
the Caputo fractional derivative. \\

\vskip 3mm
\noindent
{\large \bf Kilbas-Marzan Theorem.}
{\it
The Cauchy-type problem (\ref{E1}) and (\ref{E2}) and 
the nonlinear Volterra integral equation (\ref{E3}) 
are equivalent in the sense that, if  $x(t) \in C [0,t_f]$
satisfies one of these relations, then it also satisfies the other. }

\vskip 3mm
{\bf Proof.}
In \cite{KM1,KM2} (see also \cite{KST}, Theorem 3.24.) 
this theorem is proved by assuming that a function $G[t,x]$ 
for any $x \in W \subset \mathbb{R}$
belong to $C_{\gamma} (0,t_f)$ with $0 \le \gamma <1$, $\gamma <\alpha$.
Here $C_{\gamma} (0,t_f)$ is the weighted space 
of functions $f[t]$ given on $(0,t_f]$, such that  
$t^{\gamma} f[t] \in C(0,t_f)$. 
This ends the proof. $\ \ \ \Box$ \\

%%%%%%%%%%%%%%%%%%%%%%%%%%%%%%%%%%%%%%%%%%%%%%%%

We consider the fractional differential equation of the form
\be \label{E4}
_0^CD^{\alpha}_t x(t)+ 
K \, G[x(t)] \sum^{\infty}_{k=1} \delta \Bigl(\frac{t}{T}- k \Bigr)=0, 
\quad (1 <\alpha < 2) ,
\ee
where $ _0^CD^{\alpha}_t$ is the Caputo fractional derivative, 
with the initial conditions
\be \label{E5}
x(0)=x_0 , \quad (D^1x)(0)=p_0 .
\ee
Using $p(t)=D^1_t x(t)$, equation (\ref{E4}) can be rewritten 
in the Hamilton form. \\

\vskip 3mm
\noindent
{\large \bf Proposition 5.}
{\it The Cauchy-type problem for the fractional differential equations 
\be
D^1_t x(t) =p(t) ,
\ee
\be 
_0^CD^{\alpha-1}_t p(t)=- 
K \, G[x(t)] \sum^{\infty}_{k=1} \delta \Bigl(\frac{t}{T}- k \Bigr) , 
\quad (1 <\alpha < 2) ,
\ee
with the initial conditions
\be 
x(0)=x_0 , \quad p(0)=p_0 
\ee
is equivalent to the discrete map equations }
\be \label{E9}
x_{n+1}=x_0+p_0(n+1)T - 
\frac{KT^{\alpha}}{\Gamma(\alpha)} \sum^{n}_{k=1} \, (n+1-k)^{\alpha-1}  G[x_k] ,
\ee
\be \label{E10}
p_{n+1}=p_0 - 
\frac{KT^{\alpha-1}}{\Gamma(\alpha-1)} \sum^{n}_{k=1} \, (n+1-k)^{\alpha-2}  G[x_k] .
\ee

\vskip 3mm
{\bf Proof.}
We use the Kilbas-Marzan theorem with the function
\[ G[t,x(t)]= - K G[x(t)] \sum^{\infty}_{k=1} \delta \Bigl(\frac{t}{T}-k \Bigr) . \]
The Cauchy-type problem (\ref{E4}) and (\ref{E5}) is equivalent to
the Volterra integral equation of second kind 
\be \label{E6}
x(t)=x_0+p_0t - \frac{K}{\Gamma(\alpha)} 
\sum^{\infty}_{k=1} \int^t_{0} d \tau \, (t-\tau)^{\alpha-1} \, G[x(\tau)] 
\, \delta \Bigl(\frac{t}{T}- k \Bigr),
\ee
in the space of continuously differentiable functions $x(t) \in C^1[0,t_f]$. 

If $nT <t< (n+1)T$, then equation (\ref{E6}) gives
\be \label{E7}
x(t)=x_0+p_0t - \frac{K T}{\Gamma(\alpha)} \sum^{n}_{k=1} 
\, (t-kT)^{\alpha-1} \, G[x(kT)] .
\ee
We define the momenta
\be  \label{pDx} p(t)=D^1_t x(t) . \ee
Then equations (\ref{E7})  and (\ref{pDx}) give
\be \label{E8}
p(t)=p_0 - \frac{K T}{\Gamma(\alpha-1)}  \sum^{n}_{k=1} \, (t-kT)^{\alpha-2}  G[x(kT)] ,
\quad (nT < t < (n+1)T) ,
\ee
where we use $\Gamma(\alpha)=(\alpha-1) \Gamma(\alpha-1)$.

The solution of the left side of the $(n+1)$-th kick (\ref{not1}) and (\ref{not2})
can be represented by equations (\ref{E9}) and (\ref{E10}),
where we use the condition of continuity $x(t_n+0)=x(t_n-0)$. 

This ends the proof. $\ \ \ \Box$ \\

{\bf Remark 9}. \\
Equations (\ref{E9}) and (\ref{E10}) define
a generalization of the universal map.
This map is derived from a fractional differential equation 
with Caputo derivatives without any approximations.
The main property of the suggested map 
is a long-term memory that means that their present state 
depends on all past states with a power-law form of weights. 

{\bf Remark 10}. \\
If $\alpha=2$, then equations (\ref{E9}) and (\ref{E10}) give 
the universal map of the form (\ref{E6i}) and (\ref{E7i})
that is equivalent to equations (\ref{UM}). 
As a result, the usual universal map is a 
special case of this universal map with memory.

{\bf Remark 11}. \\
By analogy with Proposition 5, it is easy to obtain the
universal map with memory from fractional equation (\ref{E4})
with $\alpha >2$.

%%%%%%%%%%%%%%%%%%%%%%%%%%%%%%%%%%%%%%%%%%%%%%%%%%%%%%%%%%%%%%%%%%%%%% 
\section{Conclusion}

The suggested discrete maps with memory are
generalizations of the universal map.
These maps describe fractional dynamics of complex physical systems.  
The suggested universal maps with memory are equivalent to the
correspondent fractional kicked differential equations.
We obtain a discrete map from fractional differential equation
by using the equivalence of the Cauchy-type problem and 
the nonlinear Volterra integral equation of second kind.
An approximation for fractional derivatives of these equations is not used.

It is important to obtain and to study discrete maps which correspond 
to the real physical systems described by the 
fractional differential equations.
Media with memory in mechanics and electrodynamics, we can consider
viscoelastic and dielectric materials as a media with memory.
We note that the dielectric susceptibility of a wide class of 
dielectric materials follows, over extended frequency ranges, 
a fractional power-law frequency dependence that is called
the "universal" response \cite{Jo2,Jrev}. 
As was proved in \cite{JPCM2008-2,JPCM2008-1}, 
the electromagnetic fields in such dielectric media 
are described by differential equations  
with fractional time derivatives.
These fractional equations for electromagnetic waves 
in dielectric media are common to a wide class of materials, 
regardless of the type of physical structure, 
chemical composition, or of the nature of the polarizing species, 
whether dipoles, electrons or ions.
We hope that it is possible to obtain the discrere maps with memory
which correspond to the real dielectric media described 
by the fractional differential equations.

%%%%%%%%%%%%%%%%%%%%%%%%%%%%%%%%%%%%%%%%%%%%%%%%%%%%%%%%%%%%%%%%%%%%%%%%%%%%%%

\section*{Acknowledgments}

This work was supported by the Office of Naval Research, 
Grant No. N00014-02-1-0056 and Rosnauka No.02.740.11.0244.

%%%%%%%%%%%%%%%%%%%%%%%%%%%%%%%%%%%%%%%%%%%%%%%%%%%%%%%%%%%%%%%%%%%%%%

%%%%%%%%%%%%%%%%%%%%%%%%%%%%%%%%%%%%%%%%%%%%%%%%%%%%%%%%%%%%%%%%%%%%%%%%%%%%%

\end{document}